\documentclass{llncs}
\usepackage{caption} 
\usepackage{amsmath,graphicx,wasysym}
\usepackage{amsfonts}
\usepackage{hyperref}
\usepackage{bm}
\usepackage{booktabs}
\usepackage{tabularx}
\usepackage{makecell}
\usepackage[colorinlistoftodos]{todonotes}
\usepackage{graphicx}
\usepackage{hyperref}
\usepackage{colortbl}

\definecolor{Gray}{gray}{0.9}

\begin{document}
\title{Deep Learning for Quality Control of Subcortical Brain 3D Shape Models} 
\author{The ENIGMA Consortium}
\institute{The full author list appears at the end of the paper}
\maketitle              

\begin{abstract}

We present several deep learning models for assessing the morphometric fidelity of deep grey matter region models extracted from brain MRI. We test three different convolutional neural net architectures (VGGNet, ResNet and Inception) over 2D maps of geometric features. Further, we present a novel geometry feature augmentation technique based on parametric spherical mapping. Finally, we present an approach for model decision visualization, allowing human raters to see the areas of subcortical shapes most likely to be deemed of failing quality by the machine. Our training data is comprised of 5200 subjects from the ENIGMA Schizophrenia MRI cohorts, and our test dataset contains 1500 subjects from the ENIGMA Major Depressive Disorder cohorts. Our final models reduce human rater time by 46-70\%. ResNet outperforms VGGNet and Inception for all of our predictive tasks. 


\end{abstract}
\section{Introduction}
\label{sec:intro}

Quality control (QC) has become one of the main practical bottlenecks in big-data neuroimaging. Reducing human rater time via predictive modeling and automated quality control is bound to play an increasingly important role in maintaining and hastening the pace of  scientific discovery in this field. Recently, the UK Biobank publicly released over 10,000 brain MRIs (and planning to release 90,000 more); as other biobanking initiatives scale up and follow suit, automated QC becomes crucial. 

In this paper, we investigate the viability of deep convolutional neural nets for automatically labeling deep brain regional geometry models of failing quality after their extraction from brain MR images. We compare the performance of VGGNet, ResNet and Inception architectures, investigate the robustness of probability thresholds, and visualize decisions made by the trained neural nets. Our data consists of neuroimaging cohorts from the ENIGMA Schizophrenia and Major Depressive Disorder working groups participating in the ENIGMA-Shape project \cite{ENIGMAshape}. Using ENIGMA’s shape analysis protocol and rater-labeled shapes, we train a discriminative model to separate “FAIL”(F) and “PASS”(P) cases. Features are derived from standard vertex-wise measures. 

For all seven deep brain structures considered, we are able to reduce human rater time by 46 to 70 percent in out-of-sample validation, while maintaining FAIL recall rates similar to human inter-rater reliability. Our models generalize across datasets and disease samples. Our models' decision visualization, particularly ResNet, appears to capture structural abnormalities of the poor quality data that correspond to human raters' intuition. 

With this paper, we also release to the community the feature generation code based on FreeSurfer outputs, as well as pre-trained models and code for model decision visualization.

\section{Methods}

Our goal in using deep learning (DL) for automated QC differs somewhat from most predictive modeling problems. Typical two-class discriminative solutions seek to balance misclassification rates of each class. In the case of QC, we focus primarily on correctly identifying FAIL cases, by far the smaller of the two classes \textbf{(Table \ref{table:data-overview})}. In this first effort to automate shape QC, we do not attempt to eliminate human involvement, but simply to reduce it by focusing human rater time on a smaller subsample of the data containing nearly all the failing cases. 

\subsection{MRI processing and shape features}

Our deep brain structure shape measures are computed using a previously described pipeline \cite{ShapeISBI} \cite{ShapeNature}, available via the ENIGMA Shape package. Briefly, structural MR images are parcellated into cortical and subcortical regions using FreeSurfer. Among the 19 cohorts participating in this study, FreeSurfer versions 5.1 and 5.3 were used. The binary region of interest (ROI) images are then surfaced with triangle meshes and spherically registered to a common region-specific template \cite{SphericalDemonsGutman}. This leads to a one-to-one surface correspondence across the dataset at roughly 2,500 vertices per ROI. Our ROIs include the left and right thalamus, caudate, putamen, pallidum, hippocampus, amygdala, and nucleus accumbens. Each vertex $p$ of mesh model $\mathcal{M}$ is endowed with two shape descriptors:

	Medial Thickness, $D(p)=\Vert c_p - p\Vert$, where $c_p$ is the point on the medial curve $c$ closest to $p$. 

	$LogJac(p)$, Log of the Jacobian determinant $J$ arising from the template mapping,  $J:T_{\phi(p)}  \mathcal{M}_t \to T_p \mathcal{M}$.

Since the ENIGMA surface atlas is in symmetric correspondence, i.e., the left and right shapes are vertex-wise symmetrically registered, we can combine the two hemispheres for each region for the purposes of predictive modeling. Though we assume no hemispheric bias in QC failure, we effectively double our sample. 

The vertex-wise features above are augmented with their volume-normalized counterparts: $\{D,J\}_{normed}(p) = \frac{\{D,J\}(p)}{V^{\{\frac{1}{3},\frac{2}{3}\}}}$. Given discrete area elements of the template at vertex $p$, $A_t(p)$, we estimate volume as $V = \sum\limits_{p\in vrts(\mathcal{M})}^{}3A_t(p)J(p)D(p)$. We normalize our features subject-wise by this volume estimate to control for subcortical structure size. 

\subsection{Human quality rating}
Human-rated quality control of shape models is performed following the ENIGMA-Shape QC protocol (\href{enigma.usc.edu/ongoing/enigma-shape-analysis}{enigma.usc.edu/ongoing/enigma-shape-analysis}). Briefly, raters are provided with several snapshots of each region model as well as its placement in several anatomical MR slices. A guide with examples of FAIL (QC=1) and PASS (QC=3) cases is provided to raters, with an additional category of MODERATE PASS (QC=2) suggested for inexperienced raters. With sufficient experience, the rater typically switches to the binary FAIL/PASS rating. In this work, all QC=2 cases are treated as PASS cases, consistent with ENIGMA shape studies. 



\subsection{Feature mapping to 2D images}

Because our data resides on irregular mesh vertices, we first interpolate the features from an irregular spherical mesh onto an equiangular grid. The interpolated feature maps are then treated as regular 2D images by Mercator projection. Our map is based on the medial curve-based global orientation function (see \cite{ShapeMedial2012}), which defines the latitude ($\theta$) coordinate, as well as a rotational standardization of the thickness profile $D(p)$ to normalize the longitudinal ($\phi$) coordinate. The resulting map normalizes the 2D appearance of $D(p)$, setting the poles to lie at the ends of the medial curve. In practice, the re-sampling is realized as matrix multiplication based on trilinear mesh interpolation, resulting in a 128$\times$128 image for each measure. 

\subsection{Data augmentation}

Although our raw sample of roughly 13,500 examples is exceptionally large by the standards of neuroimaging, this dataset may not be large enough to train generalizable CNNs. Standard image augmentation techniques, e.g. cropping and rotations, are inapplicable to our data. To augment our sample of spherically mapped shape features, we sample from a distribution of spherical deformations, i.e. changes in the spherical coordinates of the thickness and Jacobian features. To do this, we first sample from a uniform distribution of vector spherical harmonic coefficients $\boldsymbol{B}_{lm},\boldsymbol{C}_{lm}$, and apply a heat kernel operator \cite{SphericalDemonsGutman} to the generated field on $T \mathbb{S}^2$. Change in spherical coordinates is then defined based on the tangential projection of the vector field, as in \cite{SphericalDemonsGutman}. The width $\sigma$ of the heat kernel defines the level of smoothness of the resulting deformation, and the maximum point norm $M$ defines the magnitude. In practice, each random sampling is a composition of a large magnitude, smooth deformation ($\sigma = 10^{-1}$,  $M = 3 \times 10^{-1}$) and a smaller noisier deformation ($\sigma = 10^{-2}$, $M = 3 \times 10^{-2}$). Once the deformation is generated, it is applied to the spherical coordinates of the irregular mesh, and a new sampling matrix is generated, as above.

\subsection{Deep learning models}

We train VGGNet \cite{simonyan2014very}, ResNet \cite{he2016deep} and Inception \cite{szegedy2017inception} architectures on our data. We chose these architectures as they perform well in traditional image classification problems and are well-studied.  


\subsection{Model decision visualization}

Deep learning models tend to learn superficial statistical patterns rather than high-level global or abstract concepts \cite{jo2017measuring}. As we plan to provide a tool that both (1) classifies morphometric shapes, and (2) allows a user to visualize what the machine perceives as a 'FAIL', model decision visualization is an important part of our work. Here, we use Prediction Difference Analysis \cite{zintgraf2017visualizing}, and Grad-CAM \cite{selvaraju2016grad}  to visualize 'bad' and 'good' areas for each particular shape in question. 

\subsection{Predictive model assessment}

We use two sets of measures to evaluate the performance of our models. To assess the validity of the models' estimated 'FAIL' probabilities, we calculate the area under the ROC curve (ROC AUC). We also use two supplementary measures: FAIL-recall and FAIL-share. In describing them below, we use the following definitions. TF stands for TRUE FAIL, FF stands for FALSE FAIL, TP stands for TRUE PASS, and FP stands for FALSE PASS. Our first measure, $\textbf{F-recall} = \frac{TF}{TF + FP}$, shows the proportion of FAILS that are correctly labeled by the predictive model with given probability threshold. The second measure, $\textbf{F-share} = \frac{TF + FF}{\text{Number of observations}}$, shows the proportion of the test sample labeled as FAIL by the model. Ideal models produce minimal \textbf{F-share}, and an \textbf{F-recall} of 1 for a given set of parameters. 

\section{Experiments}

For each of the seven ROIs, we performed three experiments defined by three DL models (VGGNet-, ResNet- and Inception-like architecture). 

\subsection{Datasets}

Our experimental data from the ENIGMA working groups is described in \textbf{Table \ref{table:data-overview}}. Our predictive models were trained using 15 cohorts totaling 5218 subjects' subcortical shape models from the ENIGMA-Schizophrenia working group. For a complete overview of ENIGMA-SCZ projects and cohort details, see \cite{ENIGMASCZ}. 

\begin{table}
\resizebox{12.4cm}{!} {

\begin{tabular}{lllllllll}
\toprule
      &  FAIL \%            &    accumbens &      caudate &  hippocampus &     thalamus &      putamen &     pallidum &     amygdala \\
\midrule
Train & mean$\pm$std &  3.4$\pm$4.7 &  0.9$\pm$0.7 &  2.0$\pm$1.1 &  0.8$\pm$1.0 &  0.6$\pm$0.6 &  2.3$\pm$3.6 &  0.9$\pm$0.9 \\
      & max &         16.4 &          2.1 &          4.2 &          3.4 &          1.5 &         13.8 &          2.6 \\
      & size &        10431 &        10433 &        10436 &        10436 &        10436 &        10435 &        10436 \\
Test & mean$\pm$std &  4.7$\pm$4.5 &  1.4$\pm$1.5 &  4.9$\pm$4.8 &  1.4$\pm$1.5 &  0.4$\pm$0.8 &  1.9$\pm$2.0 &  0.8$\pm$0.9 \\
      & max &         10.5 &          3.5 &         11.4 &          3.5 &          1.6 &          3.8 &          2.1 \\
      & size &         3017 &         3018 &         3018 &         3018 &         3017 &         3018 &         3018 \\
\bottomrule
\end{tabular}
}
\\[10pt]
\caption{Overview of FAIL percentage mean, standard deviation and maximum for each site. Minimum is equal to 0 for all regions and sites except for hippocampus on train (FAIL percentage 5\%). Sample sizes for each ROI  vary slightly due to FreeSurfer segmentation failure.}
\label{table:data-overview}
\vspace{-2.5em}
\end{table}




To test our final models, we used data from 4 cohorts in the Major Depressive disorder working group (ENIGMA-MDD), totaling 1509 subjects, for final out-of-fold testing. A detailed description of the ENIGMA-MDD sites and its research objectives may be found here \cite{ENIGMAMDD}.


\subsection{Model validation}

All experiments were performed separately for each ROI. The training dataset was split into two parts referred to as 'TRAIN GRID' (90\% of train data) and 'TRAIN EVAL.' (10\% of the data). The two parts contained data from each ENIGMA-SCZ cohort, stratified by the cohort-specific portion of FAIL cases. 

Each model was trained on 'TRAIN GRID' using the original sampling matrix and 30 augmentation matrices resulting in 31x augmented train dataset. We also generated 31 instances of each mesh validation set using each sampling matrix and validated models' ROC AUC on this big validation set during the training.    

As models produce probability estimates of FAIL ($P_{FAIL}$), we studied the robustness of the probability thresholds for each model. To do so,  we selected $P_{FAIL}$ values corresponding to regularly spaced percentiles values of \textbf{F-share}, from 0.1 to 0.9 in 0.1 increments. For each such value, we examined \textbf{F-recall} the evaluation set.

Final thresholds were selected based on the lowest \textbf{F-share} on the TRAIN EVAL set, requiring that \textbf{F-recall} $\ge0.8$, a minimal estimate of inter-rater reliability. It is important to stress that while we used sample distribution information in selecting a threshold, the final out-of-sample prediction is made on an individual basis for each mesh. 

\section{Results}

\begin{figure}[!htb]
  \centering
  \begin{minipage}[b]{0.49\textwidth}
    \includegraphics[width=\textwidth]{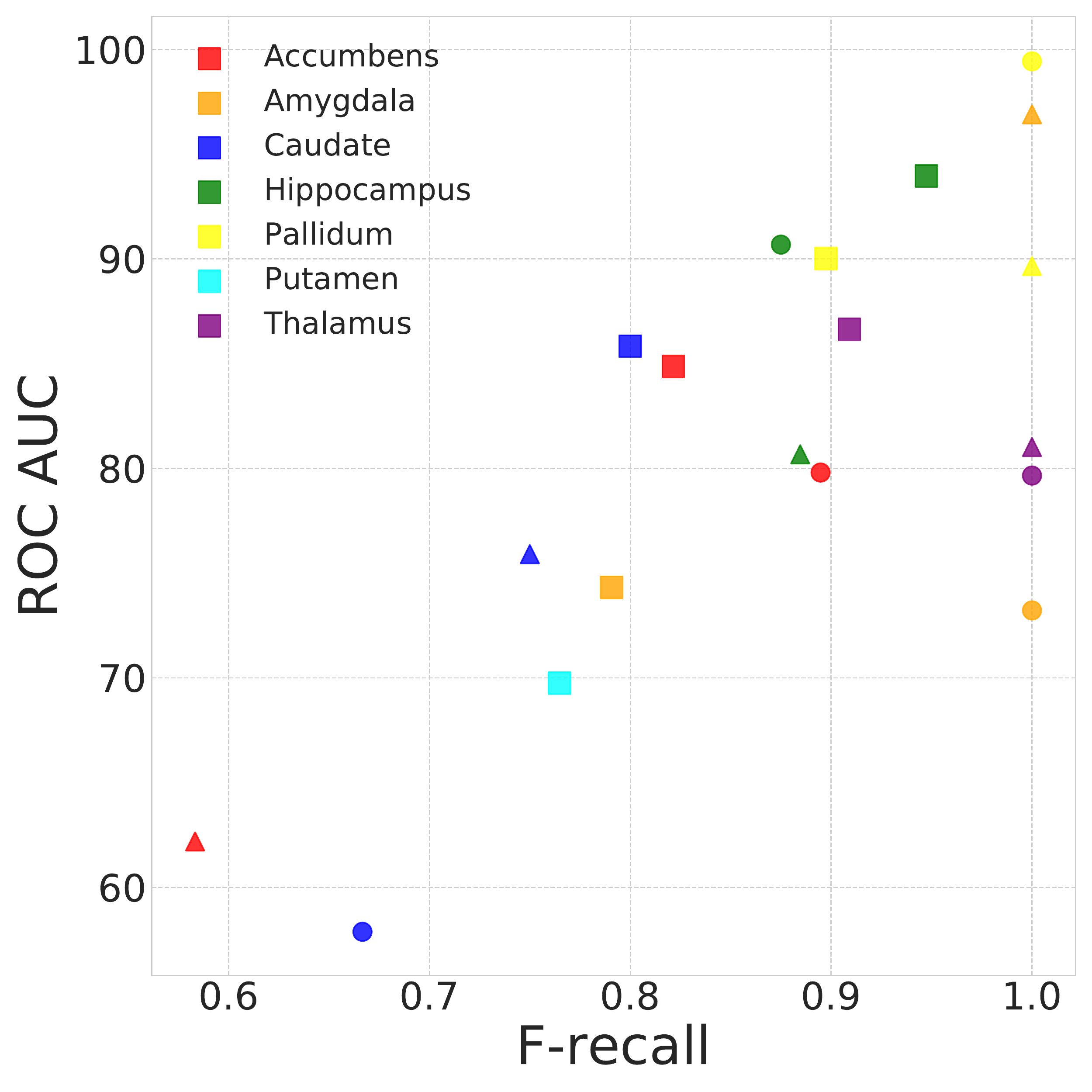}
    \end{minipage}
  \hfill
  \begin{minipage}[b]{0.49\textwidth}
    \includegraphics[width=\textwidth]{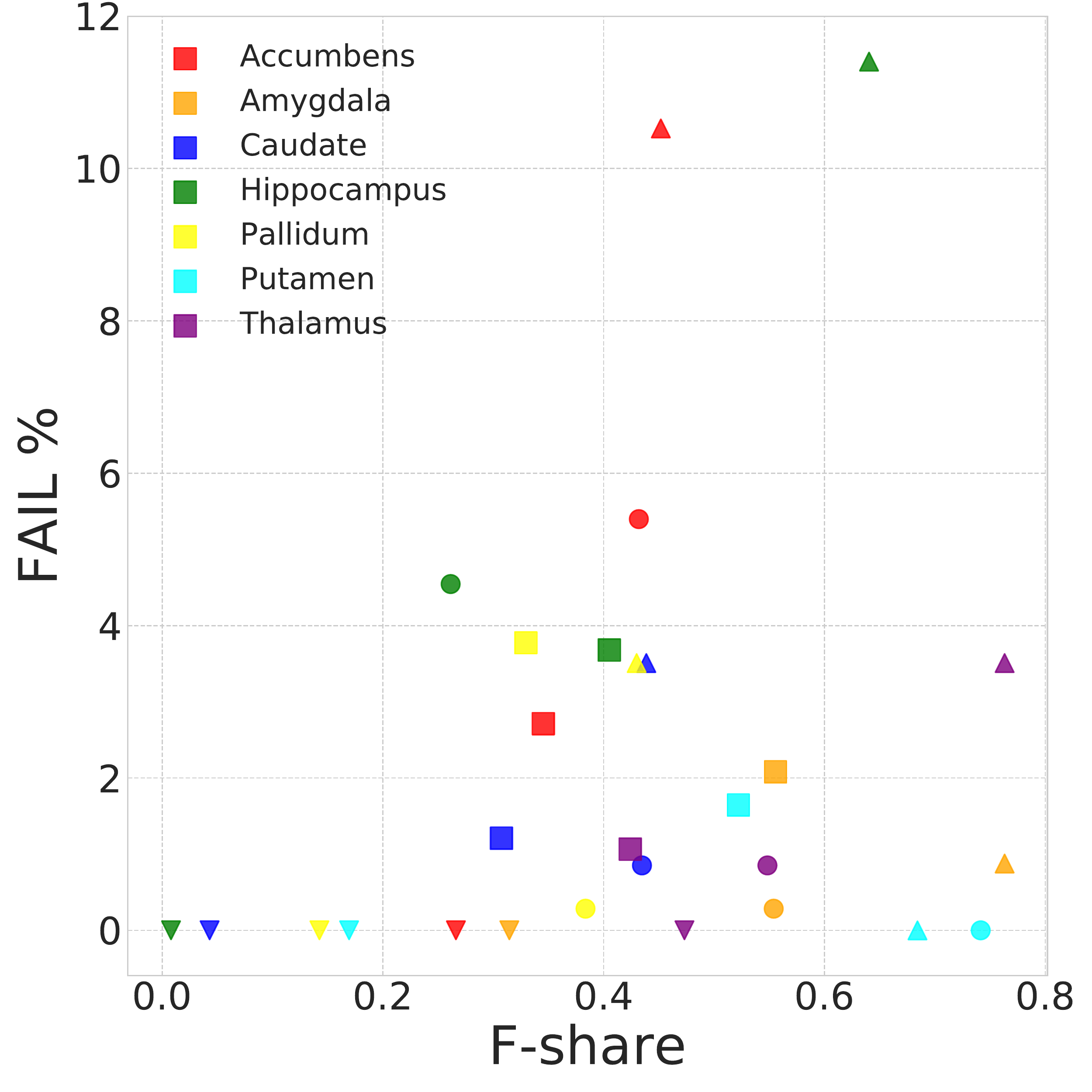}
    \end{minipage}
  \caption 
   { \label{fig:scatters} 
Scatter plots of F-recall vs. ROC AUC on test datasets and F-share vs. proportion of predicted FAIL cases on test datasets (F-share). 
\textbf{Left:} F-recall vs ROC AUC. \textbf{Right:} Fail F-share vs FAIL percentage. F-share was calculated based on thresholds from \textbf{Table~\ref{table:results-1}}.  Mark size shows the dataset size. Mark shape represents dataset (site): $\bigcirc$ - CODE-Berlin (N=176); $\Box$ - M{\"u}nster(N=1033);  $\bigtriangleup$ - Stanford (N=105); $\bigtriangledown$ - Houston(N=195)}. 
  
\end{figure}

Trained models were deliberately set to use a loose threshold for FAIL detection, predicting 0.2-0.5 of observations as FAILs in the 'TRAIN EVAL' sample. These predicted FAIL observations contained 0.85-0.9 of all true FAILs, promising to reduce the human rater QC time by 50-80\%. These results largely generalized to the test samples: \textbf{Table~\ref{table:results-1}} shows our best model and the threshold performance for each ROI. When applied to the test dataset, the models indicated modest over-fitting, with the amount of human effort reduced by 46-70\%, while capturing 76-94\% of poor quality meshes. The inverse relationship between FAIL percentage and F-share (\textbf{Figure ~\ref{fig:scatters}}) may indicate model failure to learn generalizable features on а smaller number of FAIL examples. ROC AUC and F-recall performance generalize across the test sites. Since 68\% of our test dataset is comprised of the M{\"u}nster cohort, it is important that overall test performance is not skewed by it. 

\begin{table}[!htb]
\resizebox{12.4cm}{!} {
\begin{tabular}{llllllll}
\toprule
 ROI &      Model & \thead{Eval \\AUC} & \thead{Test \\AUC} &  \thead{Eval \\F-share} & \thead{Test \\F-share} & \thead{Eval \\F-recall} & \thead{Test \\F-recall} \\
\midrule
   Accumbens &  ResNet &     0.86 &      0.8 &           0.3 &         0.35 &          0.83 &          0.78 \\
       \rowcolor{Gray}
    Amygdala &  ResNet &      0.8 &     0.75 &           0.5 &         0.54 &          0.92 &           0.8 \\
     Caudate &  ResNet &      0.9 &     0.84 &           0.2 &          0.3 &          0.82 &          0.78 \\
         \rowcolor{Gray}
 Hippocampus &  ResNet &     0.85 &     0.93 &           0.3 &         0.36 &          0.81 &          0.92 \\
    Pallidum &  ResNet &     0.86 &     0.91 &           0.3 &         0.32 &          0.81 &          0.91 \\
      \rowcolor{Gray}

     Putamen &  ResNet &     0.88 &      0.7 &           0.3 &         0.52 &          0.93 &          0.76 \\
    Thalamus &  ResNet &      0.8 &     0.87 &           0.4 &         0.47 &          0.82 &          0.94 \\
\bottomrule
\end{tabular}

}
\\[10pt]
\caption{Test performance of the best models for each region. ResNet performs the best in all cases. Overall models' performance generalizes to out-of-sample test data.}
\label{table:results-1}
\end{table}


\begin{figure}[!htb]
  \centering
  \begin{minipage}[b]{0.39\textwidth}
    \includegraphics[width=\textwidth]{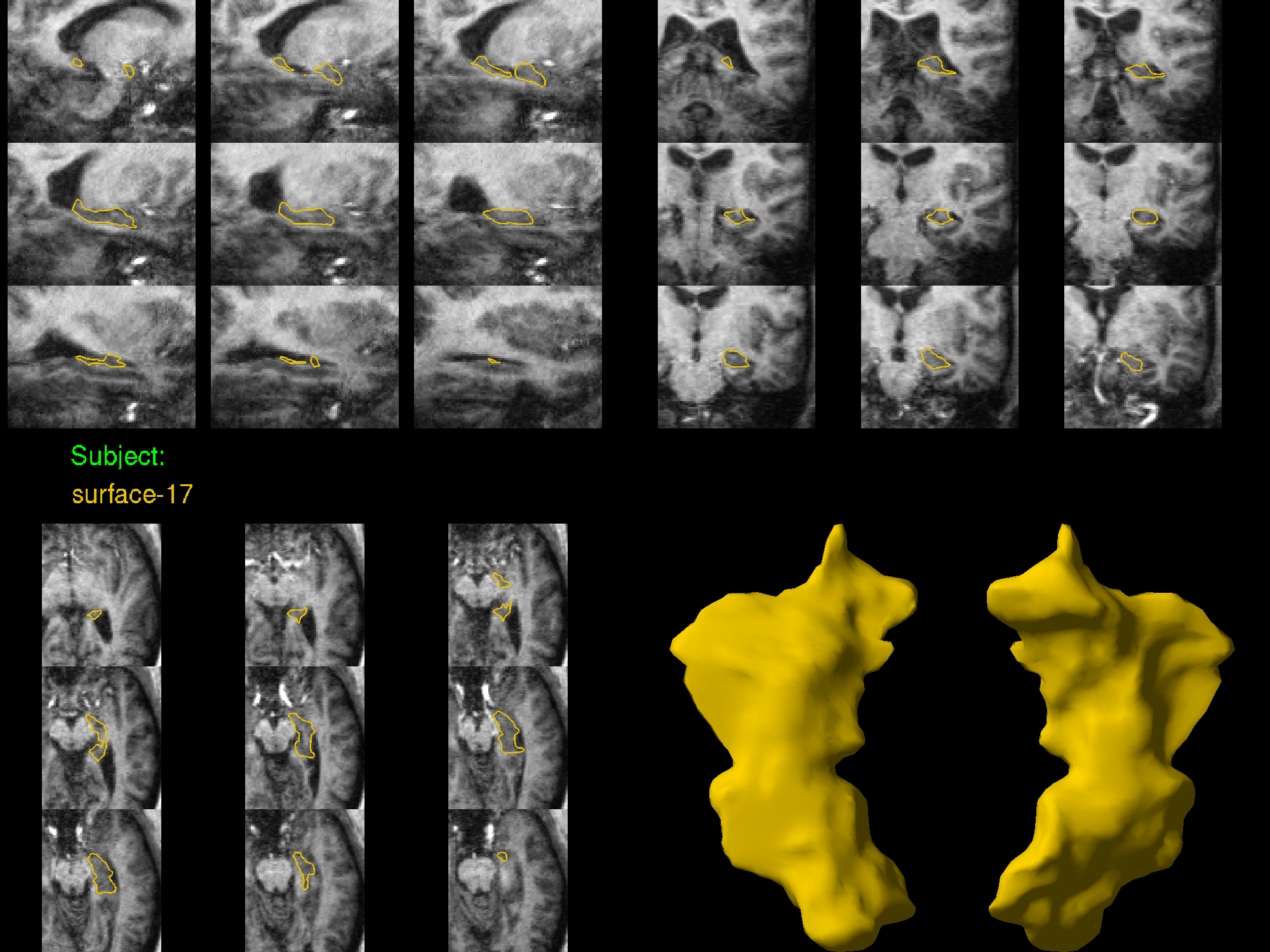}
    \end{minipage}
  \hfill
  \begin{minipage}[b]{0.6\textwidth}
    \includegraphics[width=\textwidth]{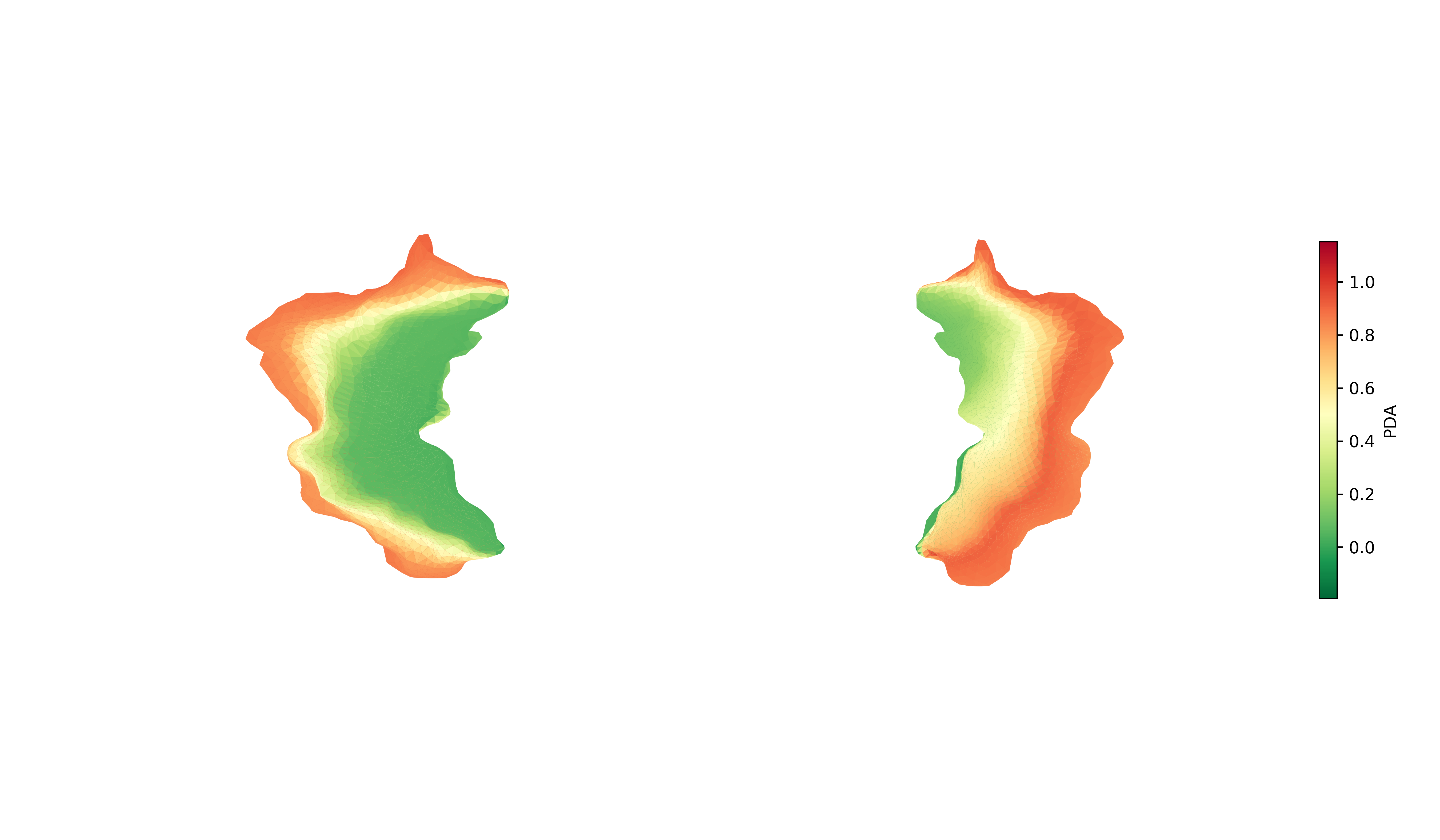}
    \end{minipage}
  \caption 
   { \label{fig:vis} 
QC report for human raters (\textbf{left}) and decision visualization example based on Grad-CAM for the ResNet model (\textbf{right}). Red colors correspond to points maximizing the model's FAIL decision in the last layer. Decision visualization corresponds to the observable deviations from underlying anatomical boundaries indicative of a "FAIL" rating according to an experienced rater.
}. 
\end{figure}

Our experiments with decision visualization (see \textbf{Fig. \ref{fig:vis}}) indicate that in most FAIL cases, the attention heat map generated by Grad-CAM corresponds to human raters' intuition while Prediction Difference Analysis tend to concentrate on local 'bumps' on shapes.

\section{Discussion and Conclusion}

We have presented potential deep learning solutions for semi-automated quality control of deep brain structure shape data. We believe this is the first DL approach for detecting end-of-the-pipeline feature failures in deep brain structure geometry. We showed that DL can robustly reduce human visual QC time by 46-70\% for large-scale analyses, for all seven regions in question, across diverse MRI datasets and populations. Qualitative analysis of models decisions shows promise as a potential training and heuristic validation tool for human raters. 

There are several limitations of our work. Our planar projection of vertex-wise features introduces space-varying distortions and boundary effects that can affect training, performance and visualization. Recently proposed spherical convolutional neural nets \cite{cohen2018spherical} may be useful to fix this issue. Second, our models' decision visualization only partly matches with human raters' intuition. In some cases, our models do not consider primary "failure" areas, as assessed by a human rater. Finally, our models are trained on purely geometrical features and do not include information on shape boundaries inside the brain. In rare cases, human raters pass shapes with atypical geometry because their boundaries look reasonable, and conversely mark normal-appearing geometry as failing due to poor a fit with the MR image. Incorporating intensity as well as geometry features will be the focus of our future work.

\bibliography{shape_refs}

\begin{thebibliography}{10}

\bibitem{ENIGMAshape}
Gutman, B., Ching, C., Andreassen, O., Schmaal, L., Veltman, D., Van~Erp, T.,
  Turner, J., Thompson, P.M.,  et~al.:
\newblock Harmonized large-scale anatomical shape analysis: Mapping subcortical
  differences across the {ENIGMA} {B}ipolar, {S}chizophrenia, and {M}ajor
  {D}epression working groups.
\newblock Biological Psychiatry \textbf{81}(10) (2017)  S308

\bibitem{ShapeISBI}
Gutman, B.A., Jahanshad, N., Ching, C.R., Wang, Y., Kochunov, P.V., Nichols,
  T.E., Thompson, P.M.:
\newblock Medial demons registration localizes the degree of genetic influence
  over subcortical shape variability: An n= 1480 meta-analysis.
\newblock In: Biomedical Imaging (ISBI), 2015 IEEE 12th International Symposium
  on, IEEE  1402--1406

\bibitem{ShapeNature}
Roshchupkin*, G.V., Gutman*, B.A.,  et~al.:
\newblock Heritability of the shape of subcortical brain structures in the
  general population.
\newblock Nature Communications \textbf{7} (2016)  13738

\bibitem{SphericalDemonsGutman}
Gutman, B.A., Madsen, S.K., Toga, A.W., Thompson, P.M.:
\newblock 24.
\newblock In: A Family of Fast Spherical Registration Algorithms for Cortical
  Shapes. Volume 8159 of Lecture Notes in Computer Science. Springer
  International Publishing (2013)  246--257

\bibitem{ShapeMedial2012}
Gutman, B.A., Yalin, W., Rajagopalan, P., Toga, A.W., Thompson, P.M.:
\newblock Shape matching with medial curves and 1-d group-wise registration.
\newblock In: Biomedical Imaging (ISBI), 2012 9th IEEE International Symposium
  on.  716--719

\bibitem{simonyan2014very}
Simonyan, K., Zisserman, A.:
\newblock Very deep convolutional networks for large-scale image recognition.
\newblock arXiv preprint arXiv:1409.1556 (2014)

\bibitem{he2016deep}
He, K., Zhang, X., Ren, S., Sun, J.:
\newblock Deep residual learning for image recognition.
\newblock In: Proceedings of the IEEE conference on computer vision and pattern
  recognition. (2016)  770--778

\bibitem{szegedy2017inception}
Szegedy, C., Ioffe, S., Vanhoucke, V., Alemi, A.A.:
\newblock Inception-v4, inception-resnet and the impact of residual connections
  on learning.
\newblock In: AAAI. Volume~4. (2017) ~12

\bibitem{jo2017measuring}
Jo, J., Bengio, Y.:
\newblock Measuring the tendency of cnns to learn surface statistical
  regularities.
\newblock arXiv preprint arXiv:1711.11561 (2017)

\bibitem{zintgraf2017visualizing}
Zintgraf, L.M., Cohen, T.S., Adel, T., Welling, M.:
\newblock Visualizing deep neural network decisions: Prediction difference
  analysis.
\newblock arXiv preprint arXiv:1702.04595 (2017)

\bibitem{selvaraju2016grad}
Selvaraju, R.R., Cogswell, M., Das, A., Vedantam, R., Parikh, D., Batra, D.:
\newblock Grad-cam: Visual explanations from deep networks via gradient-based
  localization.
\newblock See https://arxiv. org/abs/1610.02391 v3 \textbf{7}(8) (2016)

\bibitem{ENIGMASCZ}
van Erp, T.G.M., Hibar, D.P.,  et~al.:
\newblock Subcortical brain volume abnormalities in 2028 individuals with
  schizophrenia and 2540 healthy controls via the {ENIGMA} consortium.
\newblock Molecular psychiatry (2015)

\bibitem{ENIGMAMDD}
Schmaal, L., Hibar, D.P.,  et~al.:
\newblock Cortical abnormalities in adults and adolescents with major
  depression based on brain scans from 20 cohorts worldwide in the enigma major
  depressive disorder working group.
\newblock Mol Psychiatry \textbf{22}(6) (2017)  900--909

\bibitem{cohen2018spherical}
Cohen, T.S., Geiger, M., Koehler, J., Welling, M.:
\newblock Spherical cnns.
\newblock arXiv preprint arXiv:1801.10130 (2018)

\end{thebibliography}
\bibliographystyle{splncs}


\title{Full author list}

\author{Dmitry Petrov\inst{1, 2} \and Boris A. Gutman\inst{2,3} \and Egor Kuznetsov \inst{45} \and Theo G.M. van Erp\inst{4} \and  Jessica A. Turner\inst{6} \and Lianne Schmaal\inst{24,25} \and Dick Veltman\inst{25} \and Lei Wang\inst{5} \and  Kathryn Alpert\inst{5} \and  Dmitry Isaev\inst{1} \and  Artemis Zavaliangos-Petropulu\inst{1} \and Christopher R.K. Ching\inst{1} \and  Vince Calhoun\inst{40} \and  David Glahn\inst{7} \and  Theodore D. Satterthwaite\inst{8} \and  Ole Andreas Andreassen\inst{9} \and  Stefan Borgwardt\inst{10} \and  Fleur Howells\inst{11} \and Nynke Groenewold\inst{11} \and  Aristotle Voineskos\inst{11} \and  Joaquim Radua\inst{13,34,35,36} \and  Steven G. Potkin\inst{4} \and  Benedicto Crespo-Facorro\inst{14,38} \and Diana Tordesillas-Gutiérrez\inst{14,38} \and  Li Shen\inst{15} \and  Irina Lebedeva\inst{16} \and  Gianfranco Spalletta\inst{17} \and  Gary Donohoe\inst{18} \and  Peter Kochunov\inst{19} \and Pedro G.P. Rosa\inst{20,33} \and Anthony James\inst{21} \and Udo Dannlowski\inst{26} \and Bernhard T. Baune\inst{31} \and André Aleman\inst{32} \and Ian H. Gotlib\inst{27} \and Henrik Walter\inst{28} \and Martin Walter\inst{29,41,42} \and Jair C. Soares\inst{30} \and Stefan Ehrlich\inst{43} \and  Ruben C. Gur\inst{8} \and N. Trung Doan\inst{9} \and  Ingrid Agartz\inst{9} \and  Lars T. Westlye\inst{9,37} \and  Fabienne Harrisberger\inst{10} \and  Anita Riecher-R\"{o}ssler\inst{10} \and Anne Uhlmann\inst{11} \and  Dan J. Stein\inst{11} \and  Erin W. Dickie\inst{12} \and  Edith Pomarol-Clotet\inst{13,34} \and Paola Fuentes-Claramonte\inst{13,34} \and Erick Jorge Canales-Rodríguez\inst{13,34,39,46} \and Raymond Salvador\inst{13,34} \and  Alexander J. Huang\inst{4} \and  Roberto Roiz-Santiañez\inst{14,38} \and  Shan Cong\inst{15} \and  Alexander Tomyshev\inst{16} \and  Fabrizio Piras\inst{17} \and Daniela Vecchio\inst{17} \and Nerisa Banaj\inst{17} \and Valentina Ciullo\inst{17}  \and Elliot Hong\inst{19} \and Geraldo Busatto\inst{20,33} \and Marcus V. Zanetti\inst{20,33} \and Mauricio H. Serpa\inst{20,33} \and Simon Cervenka\inst{22} \and Sinead Kelly\inst{23} \and Dominik Grotegerd\inst{26} \and Matthew D. Sacchet\inst{27} \and Ilya M. Veer\inst{28} \and Meng Li\inst{29} \and Mon-Ju Wu\inst{30} \and Benson Irungu\inst{30} \and Esther Walton\inst{43,44} \and Paul M. Thompson\inst{1}, for the ENIGMA consortium}

\institute{Imaging Genetics Center, Stevens Institute for Neuroimaging and Informatics, University of Southern California, Marina Del Rey, CA, USA
\and
The Institute for Information Transmission Problems, Moscow, Russia
\and
Department of Biomedical Engineering, Illinois Institute of Technology, Chicago, IL, USA
\and
Department of Psychiatry and Human Behavior, University of California Irvine, Irvine, CA, USA
\and
Department of Psychiatry, Northwestern University, Chicago, IL, USA
\and
Psychology Department \& Neuroscience Institute, Georgia State University, Atlanta GA, USA
\and
Yale University School of Medicine, New Haven, CT, USA
\and
Department of Psychiatry, University of Pennsylvania School of Medicine, Philadelphia, PA, USA
\and
CoE NORMENT, KG Jebsen Centre for Psychosis Research, Division of Mental Health and Addiction, Oslo University Hospital \& Institute of Clinical Medicine, University of Oslo, Oslo, Norway
\and 
Department of Psychiatry, University of Basel, Basel, Switzerland
\and
MRC Unit on Risk \& Resilience to Mental Disorders, Department of Psychiatry and Mental Health, University of Cape Town, Cape Town, South Africa
\and
Centre for Addiction and Mental Health, Toronto, Canada
\and
FIDMAG Germanes Hospitalaries Research Foundation, Barcelona, Spain
\and 
University Hospital Marqués de Valdecilla, IDIVAL, Department of Psychiatry, School of Medicine, University of Cantabria, Santander, Spain
\and
Department of Radiology and Imaging Sciences, Indiana University School of Medicine, Indianapolis, IN, USA
\and
Mental Health Research Center, Moscow, Russia
\and
Laboratory of Neuropsychiatry, Santa Lucia Foundation IRCCS, Rome, Italy
\and 
School of Psychology, NUI Galway, Galway, Ireland
\and
Maryland Psychiatric Research Center,  University of Maryland School of Medicine, Baltimore
\and
Department of Psychiatry, Faculty of Medicine, University of S\~{a}o Paulo, S\~{a}o Paulo, Brazil
\and 
University of Oxford, Oxford, United Kingdom
\and
Centre for Psychiatry Research, Department of Clinical Neuroscience, Karolinska Institutet, Stockholm, Sweden 
\and 
Beth Israel Deaconess Medical Center, Harvard Medical School, Boston, MA, USA
\and
Orygen, The National Centre of Excellence in Youth Mental Health, Melbourne, Australia
\and 
Department of Psychiatry, VU University Medical Center, Amsterdam, The Netherlands
\and
Department of Psychiatry and Psychotherapy, University of M{\"u}nster, Germany
\and
Department of Psychology, Stanford University, Stanford, CA, USA
\and
Charit\'e – Universit\"atsmedizin Berlin, corporate member of Freie Universit\"at Berlin, Humboldt-Universit\"at zu Berlin, and Berlin Institute of Health, Department of Psychiatry and Psychotherapy CCM, Berlin, Germany
\and
Clinical Affective Neuroimaging Laboratory, Leibniz Institute for Neurobiology, Magdeburg, Germany
\and
University of Texas Health Science Center at Houston, Houston, TX, USA
\and 
Discipline of Psychiatry, Adelaide Medical School,
The University of Adelaide
\and
Interdisciplinary Center Psychopathology and Emotion regulation (ICPE), Neuroimaging Center (BCN-NIC), University Medical Center Groningen, University of Groningen, Groningen, The Netherlands
\and
Center for Interdisciplinary Research on Applied Neurosciences (NAPNA), University of São Paulo, São Paulo, Brazil
\and
CIBERSAM, Centro Investigación Biomédica en Red de Salud Mental, Barcelona, Spain
\and
Department of Clinical Neuroscience, Centre for Psychiatric Research, Karolinska Institutet, Stockholm, Sweden
\and
Department of Psychosis Studies, Institute of Psychiatry, Psychology and Neuroscience, King's College London, United Kingdom
\and
Department of Psychology, University of Oslo, Oslo, Norway
\and
CIBERSAM, Centro Investigación Biomédica en Red Salud Mental, Santander, Spain
\and
Radiology department, University Hospital Center (CHUV), Lausanne, Switzerland
\and
The Mind Research Network, Albuquerque, NM, USA
\and
Leibniz Institute for Neurobiology, Magdeburg, Germany
\and
Department of Psychiatry and Psychotherapy, University of Tübingen, Tübingen, Germany
\and
Division of Psychological and Social Medicine and Developmental Neurosciences, Faculty of Medicine, TU Dresden, Germany
\and
Psychology Department, Georgia State University, Atlanta, GA, USA
\and
Skolkovo Institute of Science and Technology, Moscow, Russia
\and 
Signal Processing Laboratory 5 (LTS5), \'{E}cole Polytechnique F\'{e}d\'{e}rale de Lausanne (EPFL), Lausanne, Switzerland
}

\maketitle

\end{document}